\documentclass[aps,prl,twocolumn,notitlepage,showpacs,floatfix,superscriptaddress,groupedaddress]{revtex4-1}

\usepackage{times,amsmath,amsfonts,amssymb,mathrsfs,graphics,graphicx,epstopdf,color,comment}
\usepackage[next]{inputenc}
\usepackage[dvips]{epsfig}
\usepackage[pdfstartview=FitH]{hyperref}
\usepackage{natbib}

\let\oldsqrt\sqrt
\def\sqrt{\mathpalette\DHLhksqrt}
\def\DHLhksqrt#1#2{%
\setbox0=\hbox{$#1\oldsqrt{#2\,}$}\dimen0=\ht0
\advance\dimen0-0.2\ht0
\setbox2=\hbox{\vrule height\ht0 depth -\dimen0}%
{\box0\lower0.4pt\box2}}

\newcommand{\ignore}[1]{}
\def\be{\begin{equation}}
\def\ee{\end{equation}}
\def\bea{\begin{eqnarray}}
\def\eea{\end{eqnarray}}
\def\no{\nonumber}

\begin{document}
\title{Quantum Renormalization Group for Ground-State Fidelity}

\author{A. Langari}
\affiliation{Department of Physics, Sharif University of Technology, Tehran 11155-9161, Iran} 

\author{A. T. Rezakhani}
\affiliation{Department of Physics, Sharif University of Technology, Tehran 11155-9161, Iran} 

\begin{abstract}
Ground-state fidelity (GSF) and quantum renormalization group theory (QRG) have proven useful tools in the study of quantum critical systems.  Here we lay out a general, unified formalism of GSF and QRG; specifically, we propose a method to calculate GSF through QRG, obviating the need for calculating or approximating ground states. This method thus enhances characterization of quantum criticality as well as scaling analysis of relevant properties with system size. We illustrate the formalism in the one-dimensional Ising model in a transverse field and the anisotropic spin-$1/2$ Heisenberg model.
\end{abstract}

\maketitle

\textit{Introduction.}---Quantum phase transitions (QPTs) take place in manybody systems when their ground state or few low-lying states of quantum manybody systems---describing the systems at zero or almost zero temperature---experience a considerable change with variation of Hamiltonian parameters \cite{sachdev:book}. The standard symmetry-breaking mechanism most often fails to capture QPTs. In fact, it is not always clear how to define a suitable local ``order parameter" signifying a symmetry breaking at a critical point; e.g., to systems exhibiting ``topological order," no local order parameter can be attributed \cite{wen-kitaev-top}. Moreover, discontinuities or singularities of the ground-state energy cannot always predict QPTs \cite{wolf-cirac:PRL}.  

Such inherent difficulties with identifying QPTs have necessitated tools that could better capture nature of quantum correlations. 
Along these lines, ``entanglement" has proved a useful signature for some QPTs \cite{osterloh-amico-et-al}. More interestingly, though, the elementary concept of the ``ground-state fidelity" (GSF) 
has recently been shown to provide another remarkably useful means in signaling QPTs \cite{zanardi-gsf:PRE}. This may be somehow natural as the ground state encodes all relevant information about a quantum system at zero temperature, hence a phase transition is expected to be identified by, e.g., a considerable difference between the ground states right before and right after a quantum critical point. This enables GSF as a fairly general order parameter for quantum critical systems (irrespective of their internal symmetries) \cite{zanardi-gsf:PRE,gsf-general-refs}, endowing as well a rich intrinsic geometric feature \cite{zanardi-gsf:PRL-1,rezakhani-et-al:PRA}. 

Alternatively, ``quantum renormalization group" (QRG), a variant of RG at zero temperature \cite{wilson}, puts forward a tractable recipe for studying critical behavior of a variety of quantum manybody systems, especially in one dimension \cite{pfeuty,qrg,note-comparison}. QRG essentially hinges on a coarse-graining procedure under transformation of Hamiltonian parameters, to weed out irrelevant short-distance information while retaining original large-scale picture after rescaling length. This formalism has recently been employed successfully to find critical properties of a variety of quantum manybody systems \cite{langari,kargarian1,jafari3}. 

Despite the utility of GSF, in practice its applicability is largely restricted to systems for which one can somehow compute ground state or an approximation thereof---which is a demanding task. To overcome this issue with computation of GSF, approaches based on, e.g., tensor network \cite{zhou-vidal:PRL} and Monte Carlo algorithms \cite{alet:PRL} have recently been employed. 

Here we observe that QRG can also offer a powerful alternative approach to computing GSF. Specifically, we aim to combine the GSF and QRG formalisms into a unified picture for identification of QPTs, obviating the need for the knowledge of ground state. Our formalism is fairly general and in principle can be applied to a broad class of manybody systems which are amenable to QRG formalism. To illustrate the framework, we elaborate it within two examples: (i) the Ising model in transverse field (ITF) and (ii) the anisotropic (XXZ) Heisenberg model. For specificity, in the following we adopt the Kadanoff RG recipe \cite{wilson}, although the formalism is applicable to other RG schemes as well.

\textit{Formalism.}---Consider a quantum system of $N$ spins (each with the Hilbert space $\mathcal{H}_s$ of dimension $s$), defined on a Hilbert space $\mathcal{H}^{N}_{s}\equiv \mathcal{H}_s^{\otimes N}$, with the Hamiltonian $H(\mathbf{x})$, where $\mathbf{x}$ (for simplicity taken to 
be a single parameter) is a coupling constant. In the renormalization procedure, 
the original model Hamiltonian $H$ is replaced with an effective or renormalized Hamiltonian $\widehat{H}$ 
at the cost of renormalizing coupling constants \cite{qrg,pfeuty,langari}. 
As a result, the original Hilbert space $\mathcal{H}^{N}_{s}$ is also mapped into a renormalized 
Hilbert space $\widehat{\mathcal{H}}$ encompassing only the effective degrees of freedom. 
Integrating out less important degrees of freedom gives rise to a flow in the coupling constant space. We can define an embedding 
operator $T(\mathbf{x}):~\widehat{\mathcal{H}}\to \mathcal{H}$ to 
represent this step: $T (\mathbf{x})|\Phi^{(1)}_0(\mathbf{x})\rangle=|\Phi_0(\mathbf{x})\rangle$, 
where $|\Phi_0\rangle$ and $|\Phi^{(1)}_0\rangle$ are the ground states of $H$ and $\widehat{H}\equiv T^{\dag}HT$, respectively. $T$ is usually constructed as follows: divide the system lattice into blocks of a given size, say, $m$;  considering the original Hamiltonian, attribute a Hamiltonian $h_I^B$ to each block $I$; diagonalize $h^B_I$ to find eigenvectors $\{|\phi_{i}\rangle_I\}_{i=1}^{\hat{s}}$ corresponding to first $\hat{s}$ eigenvalues to form $T_I=\sum_{i=1}^{\hat{s}}|\phi_i\rangle_I\langle \widehat{\phi}_i|$, where $\{|\widehat{\phi}_i\rangle_I\}_{i=1}^{\hat{s}}$ represent new block degrees of freedom constituting $\widehat{\mathcal{H}}_{\hat{s}}$ (and overall $\widehat{\mathcal{H}}^{N/m}_{\hat{s}}$); and, finally define the global embedding operator as $T=\otimes_{I=1}^{N/m}T_I$.   

We now recall the definition of ``fidelity" $f$, for a system of size $N<\infty$, associated to the ground states $|\Phi_0(\mathbf{x}_{\pm})\rangle$ as
\begin{eqnarray}
f &\equiv& f(\mathbf{x}, \delta; N) =
\langle \Phi_0(\mathbf{x}_{-})|\Phi_0(\mathbf{x}_{+})\rangle \label{f1}\\
& = & \langle \Phi^{(1)}_0(\mathbf{x}_{-})|T^{\dagger}(\mathbf{x}_{-}) 
T(\mathbf{x}_{+})
|\Phi^{(1)}_0(\mathbf{x}_{+})\rangle,
\label{f5}
\end{eqnarray}
where $\mathbf{x}_{\pm}=\mathbf{x} \pm \delta$, and $\delta$ represents a small variation of $\mathbf{x}$---dropping its customary absolute value for now. The group property of the renormalization procedure ensures that $|\Phi^{(1)}_0(\mathbf{x})\rangle=|\Phi_0(\mathbf{x}^{(1)})\rangle$, where $\mathbf{x}^{(1)}\equiv\mathbf{x}^{(1)}(\mathbf{x})$ is the renormalized coupling. Hence, the fidelity can be written in terms of the renormalized coupling as $f=\langle \Phi_0(\mathbf{x}_{-}^{(1)})|T^{\dag}(\mathbf{x}_-)T(\mathbf{x}_+)|\Phi_0(\mathbf{x}_{+}^{(1)})\rangle$. 
In some cases, the right-hand side may be written as a function of $f^{(1)}$, leading via RG iterations to a recurrence relation of the generic form $f^{(\ell+1)}=R(f^{(\ell)})$ ($\ell\geq 0$), where $R$ is a model-dependent function, and $f^{(\ell)}$ is the GSF after $\ell$ RG iterations. Solving this equation (analytically if the model is amenable to some exact methods) and utilizing \textit{a priori} knowledge of the associated QRG fixed points $\mathbf{x}_c$ can provide useful information such as behavior of the GSF in/around a quantum critical point or how it scales with the system size. 

Equation~(\ref{f5}) indicates that a significant simplicity ensues for the cases in which $T^{\dag}(\mathbf{x}_-)T(\mathbf{x}_+)=\omega^{(0)}(\mathbf{x},\delta)\openone$ with some $\omega^{(0)}$ (hence $R$ is linear) \cite{note}; specifically, here $f=\omega^{(0)}f^{(1)}$. The RG iteration yields 
\begin{eqnarray}
f(\mathbf{x},\delta;N)=f^{(\ell)}(\mathbf{x},\delta)\prod_{j=0}^{\ell-1}\omega^{(j)}(\mathbf{x},\delta),
\label{f-formula}
\end{eqnarray}
reducing the computation of the GSF to $f^{(\ell)}$ and $\omega^{(\ell)}$s, in which $N=m^{\ell+1}$. An immediate consequence is that if $|\omega^{(\ell)}(\mathbf{x},\delta)|<1$~$\forall \ell$, then $\lim_{\ell\to \infty}f(\mathbf{x},\delta)=0$ [note $\lim_{\ell\to \infty}\equiv \lim_{N\to \infty}$], implying an ``orthogonality catastrophe"---hence QPT---for the corresponding $\mathbf{x}$. If for a continuum of $\mathbf{x}$s such a behavior persists, we have a critical line (as in the XXZ model discussed later).

It is often the case that rather than the GSF, the GSF  ``susceptibility," defined through Taylor expanding $f$ up to $O(\delta^2)$, $f(\mathbf{x},\delta;N)\approx 1-(\delta^2/2)\chi(\mathbf{x};N)$, suffices to capture quantum criticality \cite{gsf-general-refs,you-et-al:PRE}. Expanding the embedding operator and using the identity $T^{\dagger}(\mathbf{x}) T(\mathbf{x})=\openone$ yield 
$f \approx
\langle \Phi_0(\mathbf{x}^{(1)}_{-})| S(\mathbf{x}, \delta;N)| \Phi_0(\mathbf{x}^{(1)}_{+}) \rangle$, 
where $S\equiv\openone
+2 \delta T^{\dagger}\partial_{\mathbf{x}}T-2 \delta^2 \partial_{\mathbf{x}}T^{\dagger} \partial_{\mathbf{x}}T$. Thus the above RG procedure for $f$ applies to $\chi$ as well provided that we can treat $S$ appropriately.

\textit{The ITF model.}---This model on a periodic chain of $N$ sites is defined 
with the Hamiltonian 
\be
H(J,g)=-J\sum_{i=1}^{N}\sigma_{i}^{z}\sigma_{i+1}^{z}+g\sigma_{i}^{x},
\label{ham-itf} 
\ee 
where $J$ defines an energy scale, $g$ is the parameter that controls QPT, and $\sigma^{\alpha}_i$ is the Pauli matrix for site $i$. To apply QRG, the chain is divided into blocks of $m=2$ sites described by $H^{B}=\sum_{I=1}^{N/2}h_{I}^{B}$, where $h_{I}^{B}=-J(\sigma_{1,I}^{z}\sigma_{2,I}^{z}+g\sigma_{1,I}^{x})$ \cite{qrg}.
The block-block interaction Hamiltonian is also represented by $H^{BB}=-J\sum_{I=1}^{N/2}(\sigma_{2,I}^{z}\sigma_{1,I+1}^{z}+g\sigma_{2,I}^{x})$. $h_{I}^{B}$ can be diagonalized exactly, whence $T_I$ is constructed from the two lowest eigenstates as $T_I=|\phi_{1}\rangle_{I}\langle\Uparrow|
+|\phi_{2}\rangle_{I}\langle\Downarrow|$, in which $|\phi_{1}\rangle_I=A(g) |\uparrow\uparrow\rangle + B(g) |\downarrow\downarrow\rangle, |\phi_{2}\rangle_I=A(g) |\uparrow\downarrow\rangle + B(g) |\downarrow\uparrow\rangle$
are the two degenerate ground states of $h_{I}^{B}$. Here $\{|\uparrow\rangle,|\downarrow\rangle\}$ are the eigenvectors of $\sigma^x$, $\{|\Uparrow\rangle_{I},|\Downarrow \rangle_{I}\}$ represent the states of block $I$, and $A(g) = s/\sqrt{1+s^2}$, $B(g)=1/\sqrt{1+s^2}$, with $s=g+\sqrt{1+g^2}$. The global embedding operator $T=\otimes_{I=1}^{N/2}T_I $ leads to $\widehat{H}=T^{\dagger}(H^{B}+H^{BB})T$, which is akin to the original one [Eq.~(\ref{ham-itf})] modulo replacing the coupling constants with  
\bea 
\label{jpgp-itf}
J^{(1)}=J\frac{2(g+\sqrt{1+g^{2}})}{1+(g+\sqrt{1+g^{2}})^{2}}~~,~~
g^{(1)}=g^{2}.
\eea
The renormalized couplings after $\ell$ RG iterations are obtained simply from Eq.~(\ref{jpgp-itf}) upon substituting $(J,g,J^{(1)},g^{(1)})\to (J^{(\ell-1)},g^{(\ell-1)},J^{(\ell)},g^{(\ell)})$. We note that for this model the RG fixed points are $\in\{0,1,\infty\}$, from which $g_c=1$ is unstable whereas $0$ and $\infty$ are stable under the RG flow:
\begin{figure}[h]
\includegraphics[height=.8cm,width=5cm]{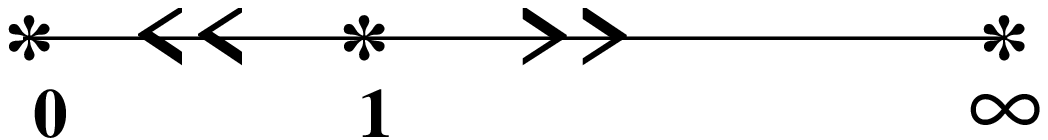}
\end{figure}

\begin{figure*}[tp]
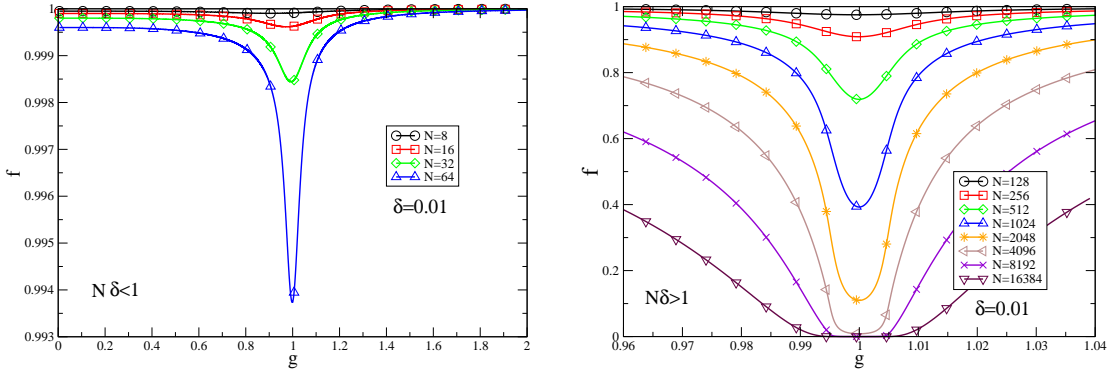

\vskip5mm
\includegraphics[width=7cm]{fig1-a.eps}
\hspace{5mm}
\includegraphics[width=7cm]{fig1-b.eps}
\caption{GSF of the ITF model vs. the field strength $g$. Left: Small-size limit $N \delta <1$. Right: Large-size limit $N \delta>1$.} 
\label{fig1}
\end{figure*}

A straightforward algebra shows that here 
$f^{(\ell)}(g,\delta)=A(g_{-}^{(\ell)}) A(g_{+}^{(\ell)})+ B(g_{-}^{(\ell)}) B(g_{+}^{(\ell)})$ 
and $\omega^{(j)}(g,\delta)=\big[A(g^{(j)}_-)A(g^{(j)}_+) + B(g^{(j)}_-)B(g^{(j)}_+)\big]^{N/2^{j+1}}$, 
from whence through Eq.~(\ref{f-formula}) one can find an analytical expression 
for $f_{\text{ITF}}(g,\delta;N)$. Figure~\ref{fig1} shows $f_{\text{ITF}}(g,\delta;N)$ for 
various values of $(\delta,N)$. A drop is seen at $g=1$, which verifies it as a quantum critical point. 
In agreement with Ref.~\cite{rams-damski:PRL}, two regimes $N\delta\lesssim1$ and $N\delta\gtrsim1$, 
corresponding respectively to the ``small-size limit" and the ``large-size limit," can be discerned. 
In the small-size limit, the GSF drops at/around $g_c=1$;  
whereas, in the large-size limit, the GSF drops to zero for a domain of $g$s around $g_c=1$ whose 
size is $\approx O(\delta)$ (Fig.~\ref{fig1} [right panel]).

In the thermodynamic limit, an interesting limit of the GSF can also be obtained. 
Equation~(\ref{jpgp-itf}) dictates three distinct regions for $g_{\pm}$ before 
applying the $\delta \rightarrow 0$ limit: (i) $g_{-}<1$ and $g_{+}<1$, (ii) $g_{-}\lesssim1$ and $g_{+}=1$, or 
$g_{-}=1$ and $g_{+}\gtrsim1$, (iii) $g_{-}>1$ and $g_{+}>1$. Thus considering the RG flow diagram of the ITF model leads to
\begin{eqnarray}
\hskip-2mm \lim_{\delta\rightarrow 0} \lim_{N \rightarrow \infty}  f_{\text{ITF}}(g, \delta; N)= 
1+\left(\frac{1+\sqrt{2}}{\sqrt{4+2\sqrt{2}}} - 1\right)\delta_{g1}, 
\label{f-limit}
\end{eqnarray}
(where $\delta_{g1}$ is the Kronecker delta function) which seems to be consistent with 
the small-size limit (Fig.~\ref{fig1} [left panel]). Note that this sharp drop of the 
GSF is a signature that $g_c=1$ is a quantum critical point, consistent with what QRG suggests. 

\ignore{
Now, the $S$ operator is given in the following form
in terms of block embedding operators:
\bea
S &=&\openone
+2 \delta \sum_I^{N/m} 
T_I^{\dagger} \partial_{g}T_I -2 \delta^2 \Big[ \sum_I^{N/m} \partial_{g}T_I^{\dagger} \partial_{g}T_I \nonumber\\
&&+\sum_{J\neq I}^{N/m}\sum_I^{N/m} \partial_{g}T_J^{\dagger} T_J
\partial_{g}T_I T_I^{\dagger}\Big].
\label{S-block}
\eea
The ITF model has this simplifying property that $T_I^{\dagger} \partial_{g}T_I=0$. In more general cases, such a property holds if, for example, the states building  the block embedding operator $T_I$ pertain to different subspaces with different quantum labels---i.e., associated to distinct symmetries. 
We obtain $\partial_{g}T_I^{\dagger}(g) \partial_{g}T_I(g)=D(g) \openone_I$ and $D(g)=s^2/[(1+g^2)(1+s^2)^2]$.
}

Moreover, the drop in the GSF accordingly signals a nonanalyticity in $\chi_{\text{ITF}}$, implying that here the GSF susceptibility is also a reliable to identify the criticality of the model. In fact, expanding $f(g,\delta;N)$ up to $O(\delta^2)$ (as explained in the formalism) yields the following expression:
\ignore{
 Hence, for the renormalized system
\begin{eqnarray}
f(g, \delta; N)=[1- \delta^2 N D(g)]
\langle \Phi_0(g_{-}^{(1)})|\Phi_0(g_{+}^{(1)})\rangle.
\end{eqnarray}
We iterate the RG procedure $\ell$ times, at which $\ell$ is related to the system size $N$ through $N=2^{\ell+1}$. This yields the following exact expression for the GSF:
\bea
f_{\text{ITF}}(g, \delta; N)=f^{(\ell)} \prod_{j=1}^{\ell} [1- \delta^2 2^{-(j-1)}N D(g^{(j-1)})].
\label{itf-fidelity}
\eea
}
\begin{figure}[bp]
\vskip3mm
\includegraphics[width=7cm]{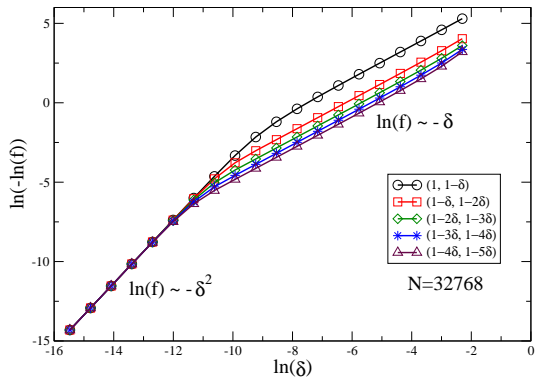}
\caption{Scaling behavior of the GSF for the ITF chain with fixed size $N=32768$:
$\ln(-\ln f )$ vs. $\ln \delta$ close
to $g_c=1$.
} 
\label{fig2}
\end{figure}
\begin{eqnarray}
\frac{\chi^{(\ell)}(g)}{2^{\ell+1}}=
\Big[ \Big(\partial_{\ell} A(g^{(\ell)})\Big)^2
+\Big(\partial_{\ell} B(g^{(\ell)}) \Big)^2 \Big] \prod_{i=1}^{\ell} g^{(i-1)}, 
\label{itfchi-n}
\end{eqnarray}
with $\partial_{\ell}\equiv \partial_{g^{(\ell)}}$, thence 
$\chi_{\text{ITF}}(g)=\chi^{(\ell)}(g)+N \sum_{j=1}^{\ell} D(g^{(j)})/2^{j-1}$, 
where $D(g)=s^2/[(1+g^2)(1+s^2)^2]$. Note that $g^{(\ell)}=g^{(\ell-1)}=g_c=1$, $\chi^{(\ell)}(1)=N^2/32$, hence $\chi_{\text{ITF}}(g_c) \sim N^2$ \cite{gsf-general-refs}. Alternatively, for second-order QPTs, it has been known that $\chi(g_c)\sim N^{2/d\nu}$, where $d$ is the dimensionality and $\nu>0$ is the critical exponent capturing the divergence of the correlation length $\xi\sim |g-g_c|^{-\nu}$ \cite{zanardi-gsf:PRL-2,rezakhani-et-al:PRA,alet-et-al:PRB}. Thus we obtain $\nu_{\text{ITF}}=1$.

\begin{figure}[bp]
\vskip4mm
\includegraphics[width=7cm]{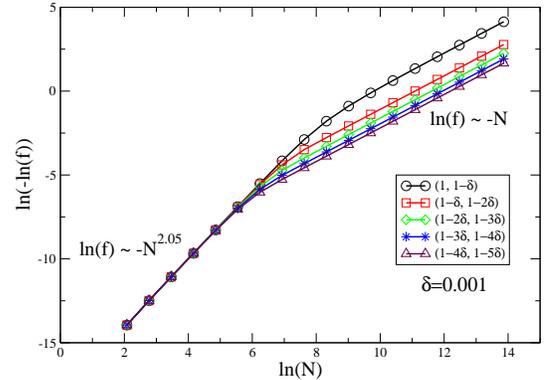}
\caption{Scaling of the GSF for the ITF chain with fixed $\delta=0.001$: 
$\ln(-\ln f)$ vs. $\ln N$ close to $g_c=1$. 
} 
\label{fig3}
\end{figure}

\ignore{
It is interesting to note that the unified formalism captures both of the small- and large-size behaviors of the GSF. However, a remark on the domain of validity of the derived behaviors is in order. Note that the embedding operator has been expanded up to $O(\delta^2)$, ignoring higher order terms thus introduces some error. Rigorously, for any small $\varepsilon$ there exists a $\delta$ such that the expression Eq.~(\ref{itf-fidelity}) is the limit of the true GSF up to an $\varepsilon$ deviation. Hence, for any QRG iteration $j$ we have $N \delta^2|D(g^{(j-1)})|/2^{j+1}< \varepsilon$. Obviously, the left hand side decreases with $j$, while for $j=1$ this yields $N \delta^2|D(g)|/4< \varepsilon$ as a constraint on $N$ and $\delta$ within $\varepsilon$ accuracy.
}


Further scaling analysis can be made for $f_{\text{ITF}}(g,\delta;N)$ with 
fixed $N$ or $\delta$. Figures~\ref{fig2} and \ref{fig3} show $\ln(-\ln f)$ close 
to $g_c=1$ vs., respectively, $\ln \delta$ for fixed $N=32768$ and $\ln N$ for fixed $\delta=0.001$, 
comparing two ground states with $g_1=1-k \delta$ and $g_2=1-(k+1)\delta$, where $k\in\{0, 1, 2, 3, 4\}$. 
In Fig.~\ref{fig2}, the $k=0$ case, labeled by $(1, 1-\delta)$, comparing the ground state at the 
quantum critical point and a state very close to it in the ferromagnetic phase, shows a behavior 
akin to the $k\neq0$ cases for small $\delta$s in the $N \delta <1$ regime; however, it shows 
a distinct behavior for $N \delta >1$, signaling that one 
of the states is exactly at the quantum critical point. The $k\neq0$ cases, comparing in fact 
two states in the ferromagnetic phase close to the quantum critical point, exhibit 
the $\ln f \sim -\delta^2$ scaling for $N\delta<1$, connected with a crossover domain 
of $N \delta \approx O(1)$ to the $\ln f \sim -\delta$ scaling for $N \delta >1$, 
in agreement with Ref.~\cite{rams-damski:PRL}.  
In contrast, Fig.~\ref{fig3} exhibits the $\ln f\sim -N^{2.05}$ scaling for $N \delta <1$, connected with a crossover domain of $N\delta\approx O(1)$ to the different $\ln f\sim -N$ scaling for $N \delta >1$, also in agreement with Ref.~\cite{rams-damski:PRL}. 
The distinct behavior of the $(1, 1-\delta)$ case hints that one of the states is 
at the quantum critical point. The scaling behavior of the GSF around $g_c=1$ in the 
paramagnetic phase ($g>1$) gives results similar to Figs.~\ref{fig2} and \ref{fig3}, 
which we did not present here.

\textit{The XXZ model.}---The anisotropic spin-$1/2$ Heisenberg model (XXZ)
on an open chain is defined with \cite{delgado-sierra:PRL-96}
\bea
H(J,\Delta)=J \sum_{i=1}^{N}h_{i,i+1}, \label{xxz-ham}
\eea
where $J>0$ is the exchange-energy coupling, $\Delta=(q+q^{-1})/2$ is the axial anisotropy given in terms of a pure phase $q$, and $h_{i,i+1}=\sigma_{i}^{x}\sigma_{i+1}^{x}+\sigma_{i}^{y}\sigma_{i+1}^{y}+ a_+(q)\sigma_{i}^{z}\sigma_{i+1}^{z} - a_-(q)(\sigma_{i}^{z}-\sigma_{i+1}^{z})$, with $a_{\pm}(q)=(q\pm q^{-1})/2$. Equation~(\ref{xxz-ham}) is different from the ordinary XXZ Hamiltonian in the boundary term $\propto\sigma_{1}^{z}-\sigma_{N}^{z}$, which is unimportant in the thermodynamic limit. 
It is known that this model is critical (gapless) for $|\Delta| \leq 1$ (critical line), 
exhibiting no long-range order \cite{sachdev:book}. 

\begin{figure}[tp]
\vskip5mm
\includegraphics[width=7cm]{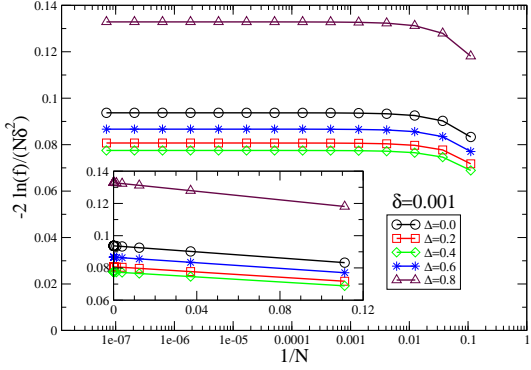}
\caption{Average GSF susceptibility vs. $1/N$ for the XXZ chain. 
In the inset the horizontal axis is in the normal scale 
(unlike the log scale of the main plot). 
The behavior in the inset is reminiscent of Fig.~1 of 
Ref.~\cite{zanardi-gsf:PRL-2} obtained through an exact diagonalization.} 
\label{fig4}
\end{figure}
The RG procedure here is implemented based on the quantum group property of the Pauli matrices \cite{delgado-sierra:PRL-96}. The Hamiltonian (\ref{xxz-ham}) is decomposed to $3$-site blocks ($m=3$), where block $I$ is comprised of sites interacting as $h^B_I=h_{i_I,i_I+1}+h_{i_I+1,i_I+2}$, and the rest of the Hamiltonian constitutes the block-block interaction. The ground state of the block Hamiltonian is doubly degenerate, represented in the 
$\sigma^z$-basis as 
\bea
|\phi_{1}\rangle&=&-C_{+}|++-\rangle+
(C_{+}+C_{-})|+-+\rangle-C_{-}|-++\rangle, \nonumber \\
\no
|\phi_{2}\rangle&=&-C_{+}|+--\rangle+
(C_{+}+C_{-})|-+-\rangle-C_{-}|--+\rangle,
\eea
where $C_{\pm}=\sqrt{q^{\pm 1}/(q+q^{-1}+4)}$. The embedding operator for block $I$ is then given by 
$T_{I}= |\phi_{1}\rangle_I \langle\Uparrow| + |\phi_{2}\rangle_I\langle \Downarrow|$,
where $\{|\Uparrow\rangle_{I},|\Downarrow \rangle_{I}\}$ denote states of block $I$.
Therefore, the renormalized Hamiltonian 
is obtained similar to Eq.~(\ref{xxz-ham}) with the following renormalized coupling constants:
\bea 
\label{qj}
q^{(1)}=q~,~ \Delta^{(1)}=\Delta~,~
J^{(1)}= J \left( \frac{q+q^{-1}+2}{2(q+q^{-1}+1)} \right)^2.
\eea

A straightforward calculation shows that here 
\begin{eqnarray}
\omega^{(0)}(\Delta,\delta)=\frac{u(\Delta,\delta)}{2\sqrt{(2+\Delta_-)(2+\Delta_+)}},
\end{eqnarray}
where 
\begin{eqnarray}
u(\Delta,\delta) &=& \sqrt{2} \Big[2\sqrt{1+\Delta_- \Delta_+ +\sqrt{(1-\Delta_-^2) (1-\Delta_+^2)}} \nonumber\\
&& + \sqrt{1+\Delta_- \Delta_+ -\sqrt{(1-\Delta_-^2) (1-\Delta_+^2)}}\Big].
\end{eqnarray}
Now through the RG formalism (especially noting that $\Delta^{(\ell)}=\Delta$), the following analytical expression is obtained for the GSF given any $(\Delta,\delta;N)$:
\begin{eqnarray}
f_{\text{XXZ}}(\Delta,\delta;N)=e^{\frac{1}{2}(N-1) \ln \omega^{(0)}(\Delta,\delta)}.
\label{f-xxz}
\end{eqnarray}
Figure~\ref{fig4} represents this scaling for some values of $\Delta$ and $\delta$. 

Interestingly, this simple and elegant relation also enables detection of the associated criticality in the XXZ model (witnessed in Refs.~\cite{zanardi-gsf:PRL-2,yang:PRA}), resolving \textit{conclusively} the doubt that the GSF might be insufficient \cite{chen-etal-KT:PRA}. Evidently we have $\omega^{(\ell)}=\omega^{(0)}$ and $|\omega^{(\ell)}|<1$ for all $\delta\neq0$ and $|\Delta|\leq1$; hence
\begin{eqnarray}
\lim_{\delta\to 0}\lim_{N\to \infty}f_{\text{XXZ}}(\Delta,\delta;N)=0,~~~\forall |\Delta|\leq 1.
\end{eqnarray}
That is, the whole $|\Delta|\leq1$ line is critical \cite{sachdev:book}. 
This is a remarkable result in that to characterize the criticality of the XXZ 
model we did not need to know the ground state or an approximation of 
that \cite{zanardi-gsf:PRL-2,yang:PRA}. We remark that the GSF 
susceptibility $\chi_{\text{XXZ}}=6(N-1)/[(1-\Delta^2)(2\Delta+4)^2]$, 
obtained through our RG approach, in contrast to the GSF only captures the criticality 
at the symmetric point $|\Delta|=1$ (including the KosterlitzÐThouless point $\Delta=1$ and 
the ferromagnetic critical point $\Delta=-1$). 

\textit{Summary.}---We have developed a viable, general quantum renormalization group 
formalism to calculate ground-state fidelity in quantum critical systems. 
Our formalism combines two powerful methods in a unified framework, enhancing characterization of criticality in quantum manybody systems. Specifically, our formalism is structured on coarse-graining a quantum system (e.g., by partitioning it into blocks) and then rescaling system length in order to eliminate short-scale or irrelevant interactions from Hamiltonian. In various cases this enables a renormalization-based recurrence relation for the fidelity, without the need to know system's ground state, an approximation thereof, or an order parameter. With this advantage, one can utilize the quantum renormalization group toolkit to boost or even simplify calculation of critical properties in systems where renormalization works sufficiently well. 

We have illustrated our formalism through two examples, the Ising model in transverse field and the anisotropic Heisenberg chain. In both models, our approach produced analytical expressions for the ground-state fidelity, resulting to the expected criticality (especially in a simpler and more conclusive way than already had been suggested for the second model).

\ignore{
Our formalism may prompt this question whether there exists a more universal picture
for studying critical properties of quantum manybody systems that can exploit advantages 
of both the renormalization group and the ground-state fidelity, an offer a general
mechanism can capture the whole spectrum of properties.
}

\textit{Acknowledgments.}---Partially supported by the Center of Excellence in
Complex Systems and Condensed Matter at Sharif University of Technology. Comments by 
N. Amiri, L. Campos Venuti, and H.-Q. Zhou are acknowledged.



\begin{thebibliography}{99}
\bibitem{note-comparison} Note that the QRG we work with is based on \textit{dynamics} not \textit{states}, thus in this respect it differs from the formalism developed in F. Verstraete \textit{et al.}, Phys. Rev. Lett. \textbf{94}, 140601 (2005).
\bibitem{note} A sufficient condition for this relation is that $\langle \phi_i(\mathbf{x}_-)|\phi_i(\mathbf{x}_+)\rangle=\omega(\mathbf{x},\delta)$, with some $i$-independent $\omega$. This might be manifestation of a selection rule associated to some symmetry.
\end{thebibliography}
\end{document}